\documentclass[12pt,preprint]{aastex} 

\def\gtorder{\mathrel{\raise.3ex\hbox{$>$}\mkern-14mu
             \lower0.6ex\hbox{$\sim$}}} 
\def\ltsima{$\; \buildrel < \over \sim \;$}
\def\simlt{\lower.5ex\hbox{\ltsima}}
\def\gtsima{$\; \buildrel > \over \sim \;$}
\def\simgt{\lower.5ex\hbox{\gtsima}} 

\begin{document} 


\title{Supernova discoveries 2010-2011: statistics and trends}


\author{Avishay Gal-Yam\altaffilmark{1},\altaffilmark{2}}
\affil{Department of Particle Physics and Astrophysics, Faculty of Physics, The Weizmann
Institute of Science, Rehovot 76100, Israel}
\email{avishay.gal-yam@weizmann.ac.il}

\author{P. A. Mazzali\altaffilmark{1}}
\affil{Astrophysics Research Institute, Liverpool John Moores University,
Liverpool, UK; INAF-Osservatorio Astronomico di Padova, 
vicolo dell'Osservatorio, 5, I-35122 Padova, Italy; Max-Planck Institut f¨ur Astrophysik, 
Karl-Schwarzschildstr. 1, D-85748 Garching, Germany}

\author{I. Manulis}
\affil{Department of Particle Physics and Astrophysics, Faculty of Physics, The Weizmann
Institute of Science, Rehovot 76100, Israel}

\and

\author{D. Bishop}
\affil{Astronomy Section, Rochester Academy of Science, Rochester NY USA}
\email{dbishop@vhdl.org}


\altaffiltext{1}{Co-chair, IAU Supernova Working Group}
\altaffiltext{2}{Kimmel Investigator}
 

\begin{abstract} 

We have inspected all supernova discoveries reported during 2010 and 2011, a total
of 538 events during 2010 and 926 events during 2011. This number includes a small number of “supernova impostors” (bright extragalactic eruptions) but not novae or events that turned out to be Galactic stars. We examine the statistics of all discovered objects, as well as those of the subset of spectroscopically-confirmed events. In these two years we see the rise of wide-field non-targeted supernova surveys to prominence, with the largest numbers of events reported by the CRTS and PTF surveys (572 and 393 events in total respectively, contributing together $74\%$ of all reported discoveries in 2011), followed by the integrated contribution of numerous amateurs (184 events). Among spectroscopically-confirmed events the PTF (393 events) leads, followed by CRTS (170 events), and amateur discoveries (144 events). Traditional galaxy-targeted surveys, such as LOSS and CHASE, maintain a strong contribution (86 and 61 events, respectively) with high spectroscopic completeness ($\sim90\%$). It is interesting to note that the community managed to provide substantial spectroscopic follow-up for relatively brighter amateur discoveries ($<{\rm m}>=16.5$\,mag), but significant less help for fainter (and much more numerous) events promptly released by the CRTS ($<{\rm m}>=18.6$\,mag). Inspecting discovery magnitude and redshift distributions we find that PS1 discoveries have similar properties ($<{\rm m}>=21.6$\,mag, $<z>=0.23$) to events found in previous seasons by cosmology-oriented projects (e.g., SDSS-II), while PTF ($<{\rm m}>=19.2$\,mag, $<z>=0.095$) and CRTS ($<{\rm m}>=18.6$\,mag, $<z>=0.049$) populate the relatively unexplored phase space of faint SNe 
($>19$\,mag) in nearby galaxies (mainly PTF), and events at $0.05< z< 0.2$ (CRTS and PTF). Examining the
specific question of reporting channels over the previous dozen years, we find that traditional reports via CBET telegrams now account only for a minority of SN discoveries. 

\end{abstract} 




\section{introduction}

The study of supernovae (SNe) has a long history, going back to pioneers like
Zwicky, Baade and Minkowski. During the decades the numbers of SN discoveries 
reported every year have increased from a handful to hundreds. With improved 
instrumentation, fainter and more distant SNe are discovered. Here, we inspect the
SN discoveries reported during the years 2010 and 2011 in detail. We also provide
a report on the use of various reporting channels during the past decade. 
Our goal is to point out general trends seen, some of which have operational
implications to the community, in particular in allocation of follow-up resources
and consideration of reporting mechanisms.

\section{Data} 

We have used the yearly summary provided by one of us (D. B.) via the 
``latest supernovae'' webpage\footnote{http://www.rochesterastronomy.org/snimages/}
to count the number of supernova discoveries during 2010-2011. As this page indexes
discoveries reported via different channels (IAUCs, CBETs, ATELs, the TOCP page) and
attempts to cross-link objects reported by different projects at 
different times, it seems optimal for this 
purpose\footnote{We have not included objects reported for the first time in refereed papers, e.g.,
those released by the PTF project in Arcavi et al. (2010) and Ofek et al. (2012; 2013),
or those reported by PS1 in Chornock et al. (2013) and Berger et al. (2012)}. 
We have included in our statistics all objects initially reported as supernovae (including “possible supernovae”). We have separately counted spectroscopically-confirmed events. In principle, we did not count the few events that turned out to be “supernova impostors”, i.e., eruptions of Luminous Blue Variables (LBVs; e.g., Van Dyk 2005, Smith et al. 2011) or members of the class of Luminous Red Novae (LRN; Kulkarni et al. 2007; for example, PTF10fqs, Kasliwal et al. 2011), and have eliminated reported events that turned out to be Galactic stars or extragalactic novae. A handful of such events (reported in 2010) may have been retained, but these should
not influence our results. Objects reported independently by two projects were counted twice (so the total discovery number is slightly less than the sum of listed discovery numbers by project). The number of such ``double-counting'' events we identified is small (34 cases in 2011). 

In general, the number of reported transients in 2011 was much larger than in 2010. In total there were 975 objects reported (551 in 2010), of these 49 ($5.0\%$) were novae, Luminous Blue Variables (LBVs), other extra-galactic variables or Galactic variable stars. These are not covered by this paper.

We list in Table 1 below the number of SN discoveries broken by projects. We have
grouped all discoveries reported by individuals into an ``amateur'' class, and
all discoveries by projects with a few SNe each into one ``other'' class. These data 
were manually harvested from the ``latest Supernovae'' webpage (accessed on Dec. 31, 2010 
and again during spring and summer 2012) and may contain errors either inherent to the webpage or 
introduced by us (if so, we apologize). Analysis of reports from 2011 was more complicated
due to the larger number of events and additional reporting channels (e.g., private webpages)
so we have tried to cross-check reported discoveries against the source of each report.
We estimate that any remaining errors that may exist are negligible, and the trends seen in this 
summary should not be affected. 

\begin{table}[h]
\begin{tabular}{lllcc}
\hline 
\hline
Project  & Total number of     & Spectroscopically      & Mean      & Mean \tabularnewline
         & SN discoveries      & confirmed events       & redshift  & magnitude [mag]\tabularnewline
         & (2010,2011,{\bf total}) & (2010,2011,{\bf total}) & (2010,2011,{\bf total}) & (2010,2011,{\bf total}) \tabularnewline
\hline 
CRTS     & 190, 382, {\bf 572}  & 67, 103, {\bf 170}     & 0.057, 0.045, {\bf 0.049} & 18.0, 18.9 {\bf 18.6} \tabularnewline 
PTF      & 88, 305, {\bf 393}   & 88, 305, {\bf 393}     & 0.086, 0.097, {\bf 0.095} & 19.3, 19.2, {\bf 19.2} \tabularnewline 
Amateurs & 82, 102, {\bf 184}   & 69, 75, {\bf 144}      & 0.019, 0.020, {\bf 0.020} & 16.7, 16.4, {\bf 16.5} \tabularnewline 
LOSS     & 50, 36, {\bf 86}     & 48, 28, {\bf 76}       & 0.024, 0.020, {\bf 0.022} & 18.2, 17.3, {\bf 17.8} \tabularnewline 
PS1      & 63, 11 {\bf 74}      & 63, 11, {\bf 74}       & 0.240, 0.169, {\bf 0.229} & 21.7, 20.8, {\bf 21.6} \tabularnewline 
CHASE    & 36, 25, {\bf 61}     & 32, 23, {\bf 55}       & 0.023, 0.016, {\bf 0.020} & 16.6, 16.8, {\bf 16.7} \tabularnewline 
MASTER   &  5, 23, {\bf 28}     & 3, 9, {\bf 12}         & 0.034, 0.159, {\bf 0.137} & 16.2, 17.6, {\bf 17.4} \tabularnewline 
ROTSE    & 13, 6, {\bf 19}      & 11, 5, {\bf 16}        & 0.051, 0.061, {\bf 0.054} & 17.9, 17.6, {\bf 17.8} \tabularnewline 
LSSSS    & 13, 6, {\bf 19}      & 9, 3, {\bf 12}         & 0.033, 0.035, {\bf 0.034} & 17.4, 16.8, {\bf 17.2} \tabularnewline 
ISSP     & 0, 19, {\bf 19}      & 0, 18, {\bf 18}        & {\bf 0.022} & {\bf 16.8} \tabularnewline
LSQ      & 0, 7, {\bf 7}        & 0, 7, {\bf 7}          & {\bf 0.061} & {\bf 18.9} \tabularnewline
Others   & 11, 6, {\bf 17}      & 9, 6, {\bf 15}         & & \tabularnewline 
\hline
\end{tabular}
\caption{Supernova discovery statistics. All explosive extragalactic 
objects reported are counted (including LBVs, LRNs and other SN impostors)
with the exception of Novae.
All discoveries reported by individuals are counted as ``amateur'' discoveries;
the ``other'' category includes all discoveries reported by projects with
a few discoveries each. Project information: Catalina Real-Time Transient Survey 
(CRTS; http://crts.caltech.edu/; Drake et al. 2009; including sub-projects MLS, SSS, SNhunt), 
Palomar Transient Factory (PTF; http://www.astro.caltech.edu/ptf/; Rau et al. 2009, 
Law et al. 2009), Pan-STARRs 1 (PS1; http://ps1sc.org/transients/), The Lick
Observatory Supernova Search (LOSS; http://astro.berkeley.edu/bait/public$\_$html/kait$\_$lwd.html;
Filippenko et al. 2001), The Chilean Automatic Supernova Search (CHASE; http://www.das.uchile.cl/proyectoCHASE/;
Pignata et al. 2009), The ROTSE Supernova Verification Project (ROTSE-RSVP; http://www.rotse.net/rsvp/),
The La Sagra Sky Survey (LSSS; http://www.minorplanets.org/OLS/LSSS.html), Mobile Astronomical System of 
TElescope Robots (MASTER; http://observ.pereplet.ru/; Lipunov et al. 2010), The Italian Supernova Search Project (ISSP; http://italiansupernovae.org/), The La-Silla Quest Variability Survey (LSQ; http://hep.yale.edu/lasillaquest). The mean magnitudes are based on values from the ``Latest Supernovae'' webpage, which 
are typically discovery magnitudes.   
}
\label{exttable}
\end{table}

\section{Discussion}

We scanned publically-accessible sources and 
have attempted to extract for as many events as we could the redshift of spectroscopically-confirmed SNe, and 
their discovery magnitudes. The data were available for the majority of events. We calculated the mean
values and present them in Table~\ref{exttable}. In the few cases where some
data were missing (mostly redshifts) we calculated the means without these data; the incompleteness appears
unlikely to change the reported numbers significantly. We did not calculate mean values for the ``other''
category, as these SNe were discovered by very different means and projects. 

Inspecting Table~\ref{exttable}, we note the following trends. 

The most prolific source of SN candidates in 2010-2011 was the CRTS survey. 
However, only $\sim30\%$ of the candidates 
discovered by this survey have been spectroscopically confirmed, many by observers that appear to be 
affiliated with the survey. It is interesting to note that the larger SN community managed to provide
substantial spectroscopic follow-up for relatively brighter amateur discoveries
($<m>=16.5$), but
did not provide significant help for slightly fainter (and much more numerous)
events promptly released by the CRTS ($<m>=18.6$). One can speculate that the
amount of large telescope time set aside specifically for spectroscopic follow-up
of SNe of unspecified origins (``public events''), in addition to 
flexible spectroscopic resources (nights during which observers can ``squeeze in'' recently
discovered SNe) are saturated by the discovery stream from CRTS.
In contrast, the LOSS survey that led this field in the previous
decade (Leaman et al. 2011, Li et al. 2011) and 
which discovers SNe of similar or fainter magnitudes to those discovered
by the CRTS, enjoyed high spectroscopic completeness ($\sim90\%$), probably due
to its dedicated long-standing follow-up program. The southern CHASE project typically discovered 
brighter events ($<m>=16.7$) and had a comparable spectroscopic completeness to that achieved for
LOSS and amateur discoveries (which also have similar magnitudes). Since PTF and PS1 only report spectroscopically-confirmed events, 
their spectroscopic completeness cannot be determined. 

\begin{figure}[h]
\centering
\includegraphics[width=1\textwidth]{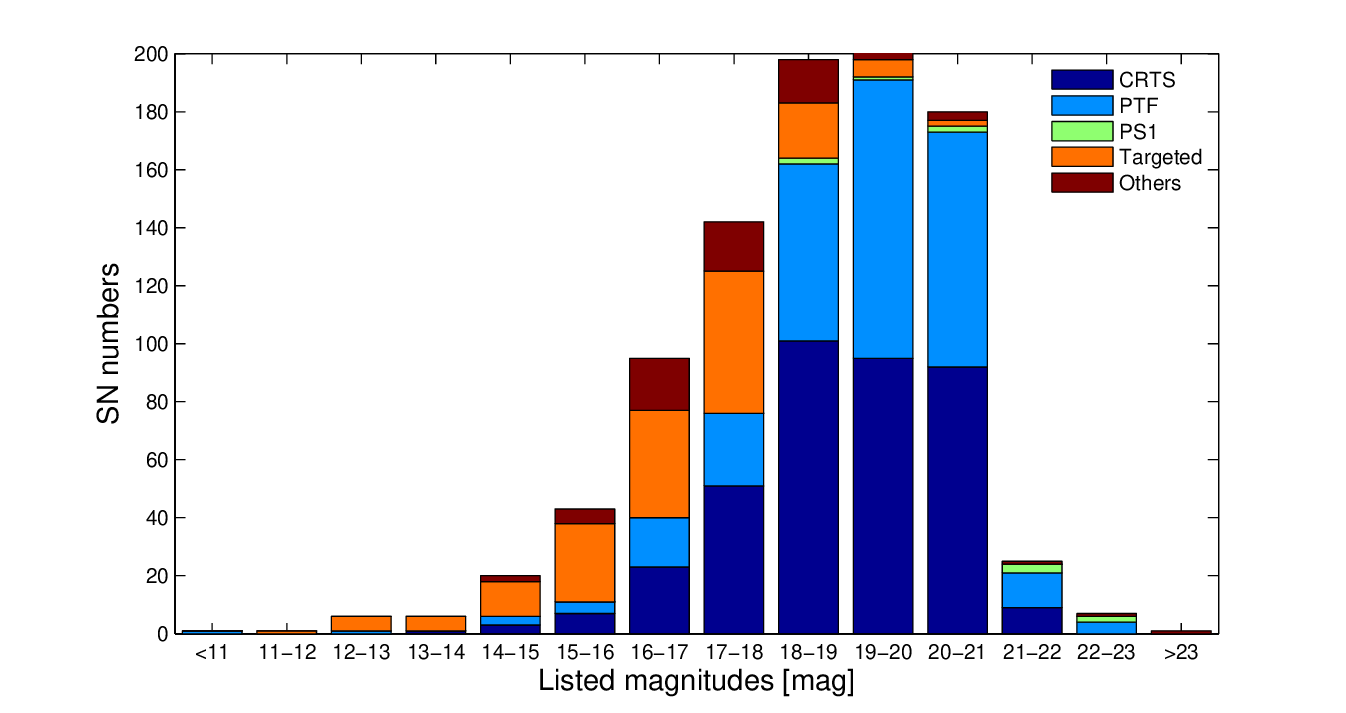}
\includegraphics[width=1\textwidth]{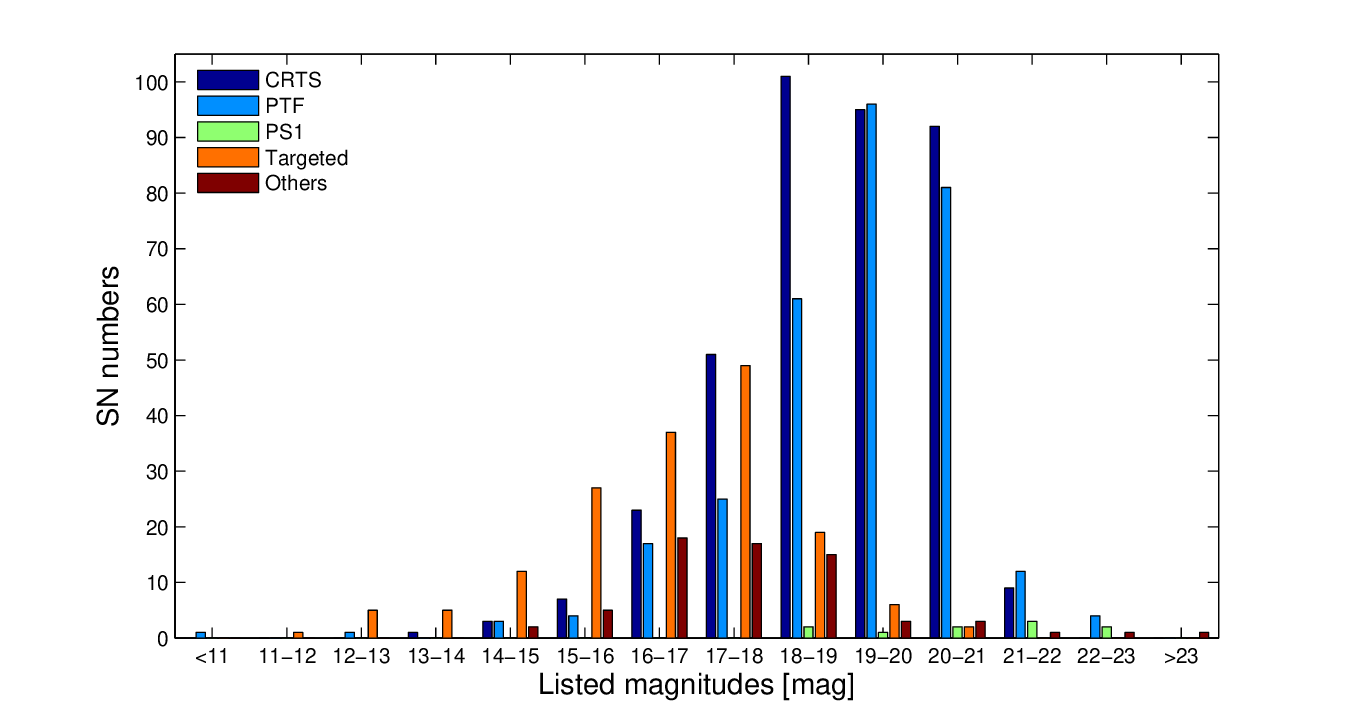}
\caption{The magnitude distribution of SNe in 2011. Events between $18-21$\,mag dominate
the numbers (top). Broken by projects, the contributions of targeted surveys (Amateurs, LOSS and 
CHASE) remain dominant at brighter magnitudes, followed (moving to fainter objects), by CRTS,
PTF and and PS1.}
\label{figmag}
\end{figure}

\begin{figure}[h]
\centering
\includegraphics[width=1\textwidth]{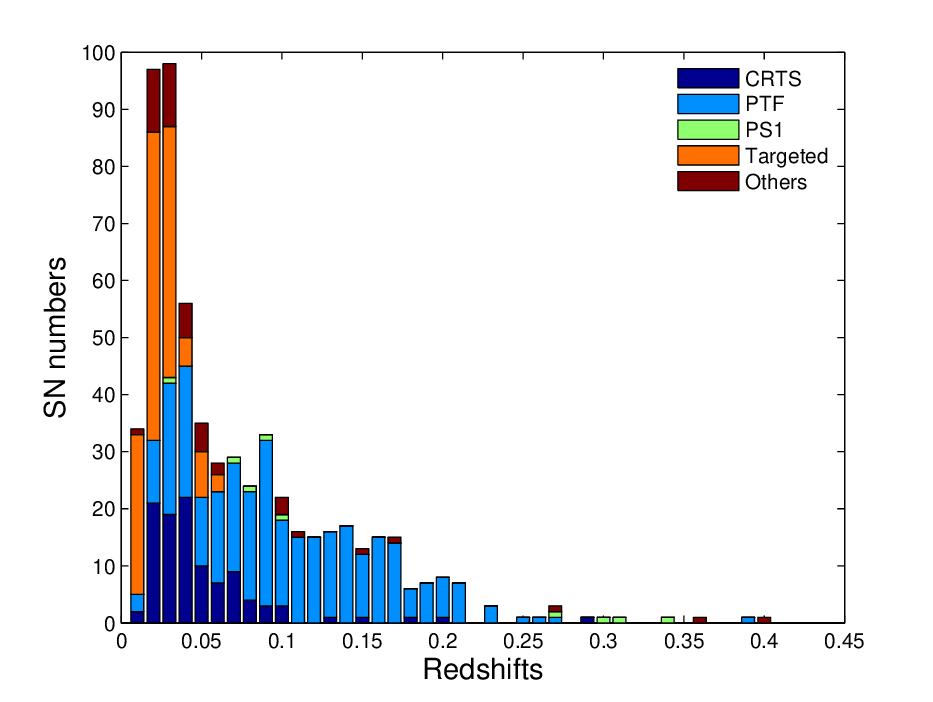}
\caption{The redshift distribution of SNe discovered during 2011. Note the strong contribution by 
Amateurs and targeted surveys at $z\sim0.02$; followed by CRTS (up to $z\sim0.1$), PTF (up to $z\sim0.2$) and PS1 ($z>0.25$).}
\label{figz}
\end{figure}

Inspecting the discovery magnitude 
and redshift distributions (Figures~\ref{figmag} and~\ref{figz}.) 
we find that PS1 discoveries have similar properties
($<m>=21.6$, $<z>\sim0.23$)
to events found in previous seasons by cosmology-oriented projects 
such as the SDSS-II supernova search (Frieman et al. 2008; Sako et al. 2008) and to a lesser degree also
ESSENCE (Miknaitis et al. 2007) and SNLS (Astier et al. 2006)\footnote{The bulk of the 
PS1 discoveries which have been announced (in 2010) are from the
PS1 Medium Deep Field survey, which covers a relatively narrow field (88 square degrees total)
to deep limiting magnitudes ($m>22$\,mag), and hence provides fainter 
targets. A small number of SNe (and other transients)
have been found (mostly in 2011) in the shallower, wider field ``3Pi Faint Galaxy Supernova
Survey'' (e.g., Valenti et al. 2010), with discovery magnitudes in the range $17 - 20$\,mag.}. 
The two other new untargeted 
surveys, PTF ($<m>=19.4$, $<z>=0.099$)
and CRTS ($<m>=18.6$, $<z>\sim0.049$) populate a relatively unexplored phase space.
PTF is finding faint SNe ($>19$\,mag) in nearby galaxies, 
while both CRTS and PTF discover events in the redshift range $0.05<z<0.2$,
previously little explored as it lies beyond the reach of the nearby targeted
surveys (LOSS and CHASE) and yet requires a wide survey to discover numerous
events (c.f., narrow and deep cosmological surveys). 

\begin{figure}[h]
\centering
\includegraphics[width=1\textwidth]{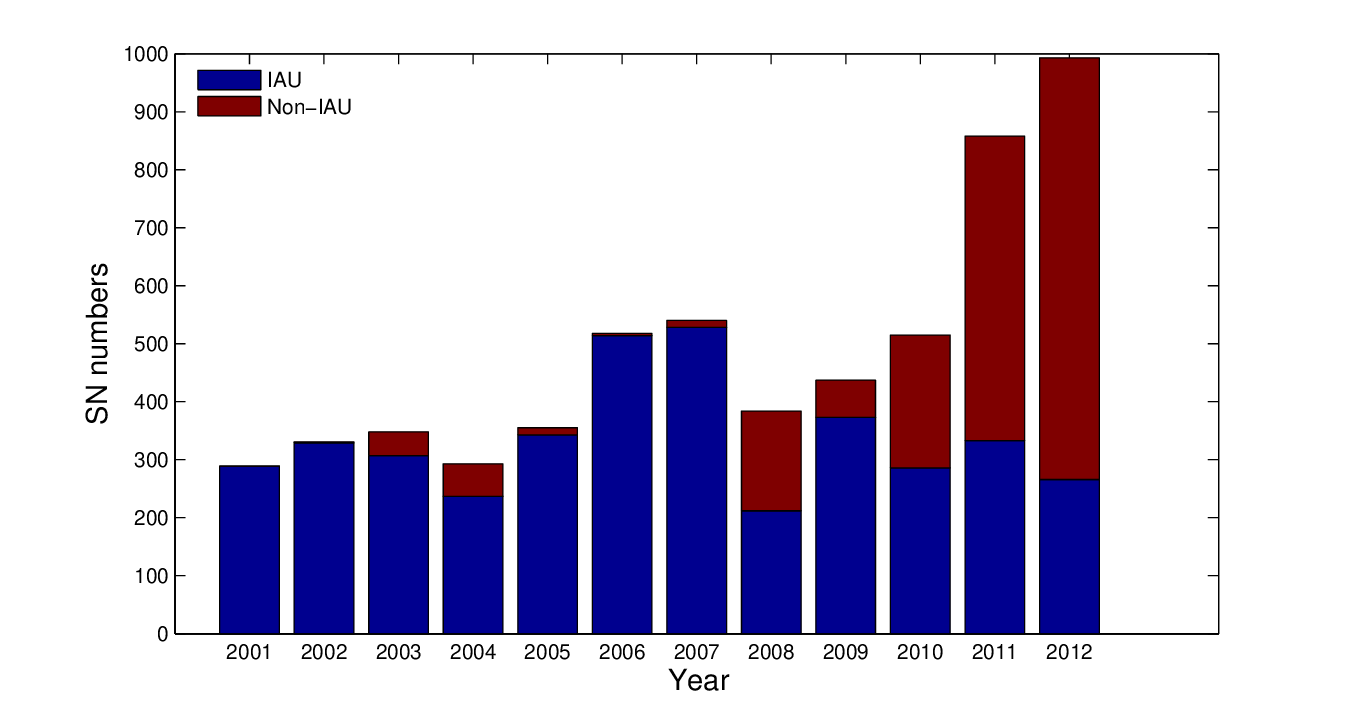}
\caption{The numbers of SNe reported to the IAU (blue) or distributed via
other channels (red). Starting in 2010, the numbers of non-IAU reports are significant,
becoming the majority starting in 2011.}
\label{figrep}
\end{figure}

In terms of reporting channels, in 2011 only 281 SNe ($30.3\%$) were reported via CBET circulars,
580 SNe ($62.6\%$) were reported in ATels, 57 SNe ($6.2\%$) were reported only via the CBAT TOCP page, and 
10 SNe ($1.1\%$) were reported using the SNHunt project webpage.
Some SNe were reported separately in both CBETs and ATels. We note that a change in CBET policy implemented in 2011 (reporting only spectroscopically-confirmed events) further complicates analysis
and record-keeping (if one includes also SNe without spectroscopic confirmation). It may seem that a 
revision should be considered by the community regarding the mode SN discoveries are reported, tracked 
and circulated. Using statistics extracted from the ``latest supernovae'' webpage for the years 2001-2012, we plot in 
Figure~\ref{figrep} the fractions of events reported to the IAU during this period. As can be seen, starting in 2010, a significant number of events are no longer reported to the IAU, and during 2011
(and more so, 2012) the non-IAU reports have become the majority.

\section{Summary}

We enumerate the reported SN discoveries during 2010-2011 and point out several statistical
trends. The impact of the new generation of untargeted surveys has been substantial
during these years, with CRTS and PTF providing the largest number of the total and spectroscopically-confirmed
events released to the community, respectively. The large number of CRTS SN candidates that have not been
spectroscopically confirmed may indicate that the SN community cannot provide follow-up to
the emerging large numbers of faint SNe. The inception and recent launch of the 
Public ESO Spectroscopic Survey of Transient Objects (PESSTO; PI Smartt\footnote{www.pessto.org})
may improve this situation. New surveys continue to provide access to new 
observational parameter space, including very faint events in nearby galaxies, and SNe in the 
relatively uncharted redshift range $0.05<z<0.2$.

\end{document}